\documentclass[12pt]{article}
\usepackage{makeidx}
\usepackage{multirow}
\usepackage{multicol}
\usepackage[dvipsnames,svgnames,table]{xcolor}
\usepackage{graphicx}
\usepackage{epstopdf}
\usepackage{ulem}
\usepackage{hyperref}
\usepackage{amsmath}
\usepackage{amssymb}
\usepackage{indentfirst}
\title{Title: Quantification of flux for non-equilibrium dynamics and thermodynamics for driving non-Michaelis-Menton Enzyme Rates}
\author{QiongLiu$^{1}$, Jin Wang$^{1,2,3*}$\\
\small{Affiliations:}\\
\small{$^{1}$State Key Laboratory of Electroanalytical Chemistry, Changchun Institute of Applied}\\
\small{Chemistry, Chinese Academy of Sciences, Changchun, Jilin, 130022, China}\\
\small{$^{2}$Department of Chemistry, Physics and Applied Mathematics, State University of New}\\
\small{York at Stony Brook, Stony Brook, New York, 11794-3400, USA}\\
\small{$^{3}$College of Physics, JilinUniversity, Changchun, Jilin, 130012, China}\\
\small{$^*$ Corresponding author. Tel: +1-631-816-5920, Fax: +1-631-632-7960.}\\
\small{E-mail address:jin.wang.1@stonybrook.edu}}
\date{}

\begin{document}
\maketitle

\section*{}

Abstract:The driving force for active physical and biological systems is determined by both the underlying landscape and the non-equilibrium curl flux. While
landscape can be quantified in the experiments by the histograms of the
collecting trajectories of the observables, the experimental flux quantification is still challenging. In this work, we studied the single molecule enzyme dynamics and observed the deviation in kinetics from the conventional Michaelis-Menton reaction rate. We identified and quantified the non-equilibrium flux as the origin of such non-Michaelis-Menton enzyme rate behavior. This is the first time of rigorous quantification of the flux for the driving force of the non-equilibrium active dynamics. We also quantified the corresponding non-equilibrium thermodynamics in terms of chemical potential and entropy production. We identified and quantified the origin of the flux, chemical potential and entropy production as the heat absorbed (energy input) in the enzyme reaction.

\section*{Main Text: }

Active physical and biological dynamical systems are everywhere around us such as atmosphere, turbulence, cells, ourselves, and even stock market etc.\cite{Jackson}. They have a common feature which is that the normal function requires the energy input from the environments. Therefore, such systems should follow the activated or non-equilibrium dynamics rather than the passive equilibrium one we normally encounter. The passive system dynamics can usually be determined by the gradient of the underlying potential landscape so that global quantifications in terms of the weight of the state and the description of local dynamics are possible. However, for the non-equilibrium activated dynamics, such physical description is not possible. Recent studies suggest that the non-equilibrium activated dynamics is determined by both the gradient of the underlying landscape and the steady state probability flux which quantifies the degree of non-equilibriumness through the activation from the energy input leading to detailed balance breaking\cite{Wang2008PNAS}. This identifies the driving force and establishes a general principle for the non-equilibrium active dynamics. Furthermore, it provides the origin of the underlying non-equilibrium thermodynamics. This idea has been applied to many physical and biological systems\cite{Wang2015AP}.While landscape can be quantified in the experiments by the histograms of the collecting trajectories of the observables, the experimental flux quantification is still challenging. In this work, we studied the single molecule enzyme dynamics and observed the deviation in kinetics from the conventional Michaelis-Menton reaction rate\cite{Xie2006, Xie2013,MichaelisMenton}. We identified and quantified the non-equilibrium flux as the origin of such non-Michaelis-Menton enzyme rate behavior. This is the first time of rigorous quantification of the flux for the driving force of the non-equilibrium active dynamics. 

Due to the technology advances, single molecule studies become possible
\cite{Moerner}. An important application of single molecule analysis was the investigation of the catalytic cycles of single enzymes\cite{Xie2006,Xie2013,Xie2007}. The single molecule features can be revealed that are hidden by the averaging process in the bulk when an ensemble of molecules is observed. The function and structure of the enzymes are closely related and small differences in the structure of the molecules can give rise to spread of kinetic rates. The heterogeneity in the kinetic rates can be revealed by single molecule measurements.

There have been growing single molecule studies of enzymatic reactions\cite{Xie2006,Xie2013}. These studies often reveal that enzyme reactions under protein fluctuations are often found to obey the classic Michaelis-Menten(MM) relation: the inverse of enzyme catalytic rate is linear to the inverse of the substrate concentration. The Michaelis-Menten form is expected to be valid as long as the detailed balance condition is preserved which is manifested by no net circulating flux between different conformations of the fluctuating enzyme\cite{Xie2007,Qian2002BPC}.

Catalysis of the oxidation of the dihydrorhodamine 6G into rhodamine 6G by the enzyme horseradish peroxidase at the single enzyme level has recently been observed at room temperature\cite{Rigler1, Rigler3, Rigler2}.
Horseradish peroxidase is an effective catalyst of the decomposition of hydrogen peroxide ($H_2O_2$)  in the presence of hydrogen donors. One particularly useful spectroscopic technique with single molecule sensitivity is fluorescence correlation spectroscopy (FCS). Rigler and collaborators used confocal FCS to monitor the catalytic rate of single horseradish peroxidase (HRP) molecules\cite{Rigler1, Rigler3}. From their measurements they concluded that the enzyme molecules displayed dynamic disorder in their enzymatic activity. HRP catalyzes the reductant of hydrogen peroxide in the presence of a reducing agent. There are many reductant agents that can be selected. HRP in the presence of hydrogen peroxide and a non-fluorescent substrate will turn over the substrate into a fluorescent product, thereby making the catalysis visible for fluorescence spectroscopy techniques. 

In this article we studied the single molecule enzyme horseradish peroxidase (HRP) reaction dynamics and quantified the flux as the non-equilibrium driving force for the deviations in kinetics from the Michaelis-Menton behavior through the correlations measured by FCS. We not only quantified the flux for driving the non-equilibrium dynamics, but also quantified the non-equilibrium thermodynamics in terms of chemical potential and entropy production as well as the time irreversibility of the underlying enzyme reaction. We quantitatively identified the origin of the flux, chemical potential and entropy production as the heat absorbed or energy input in the enzyme reaction dynamics.

\section*{Results}

\subsection*{Experimental setups}

\begin{figure}[!ht]
\centering
\includegraphics[width=0.8\textwidth]{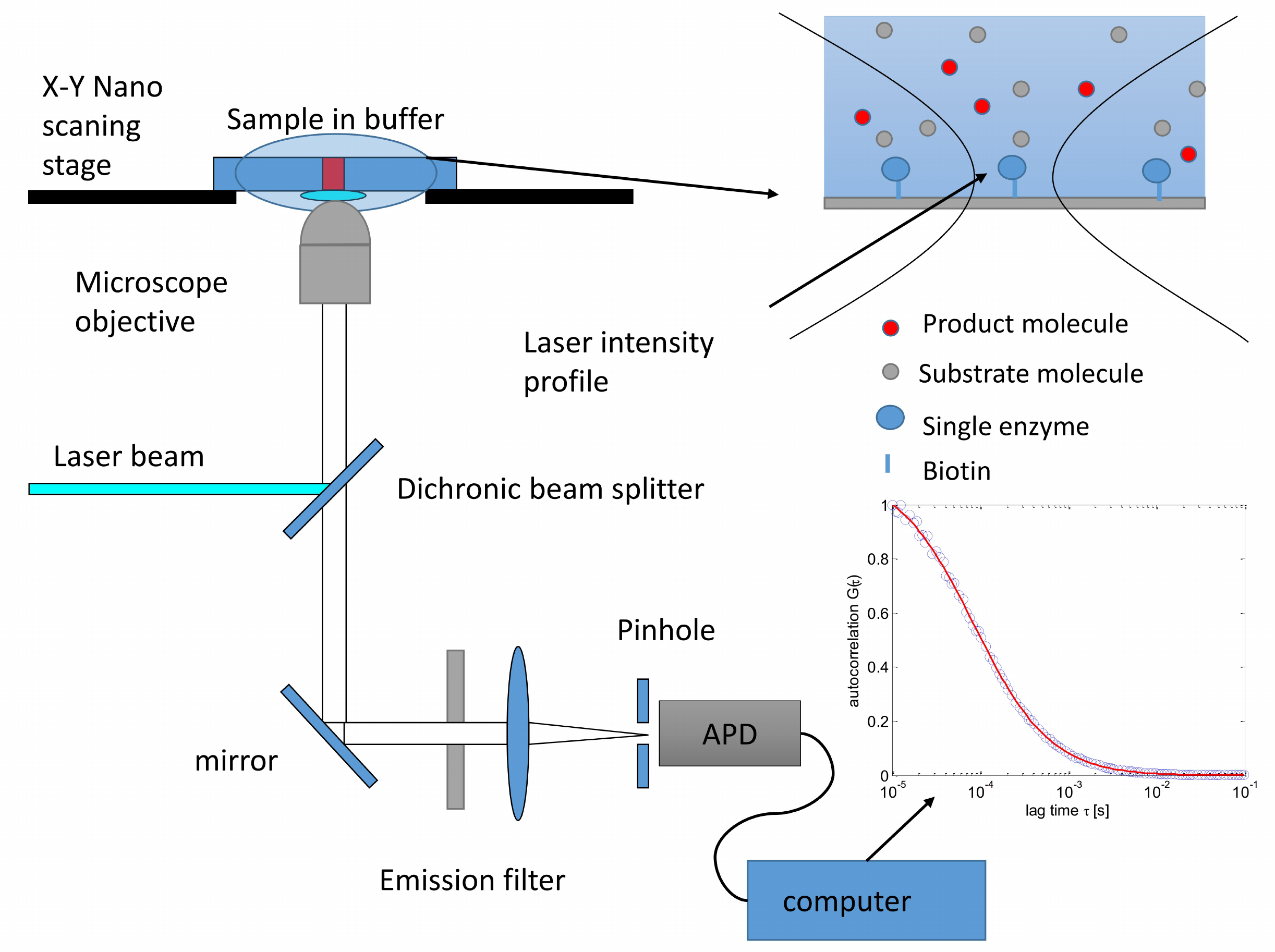}
\caption{FCS and experimental setup}\label{fig01}
\end{figure}

The FCS experimental setups consist of a high NA oil-immersion objective (1.45NA, 100*,Carl Zeiss) to excite fluorescence, an avalanche photodiode (APD) (SPCM-AQR-13, PerkinElmer) to detect fluorescence. For excitation, an ion argon laser was used with a maximum power of 8mW at 488nm.

The HRP molecules we investigated are conjugated to streptavidin. The immobilization on microscope cover slides is achieved via the very stable binding to biotin molecules, attached to the surface. The HRP was studied after immobilization on the cover slides in 100mM potassium phosphate buffer, under PH 7.0, with 1nM up to 100nM dihydrorhodamine 123, at $H_2O_2$ concentration of 25$\mu{}$M up to 1000$\mu{}$M. For graphical illustration, see Figure 1.

\subsection*{Reaction Scheme}

A graphical representation of the kinetic scheme for enzyme reaction processes we study is illustrated in Figure 2:
\begin{figure}[!ht]
\centering
\includegraphics[width=0.8\textwidth]{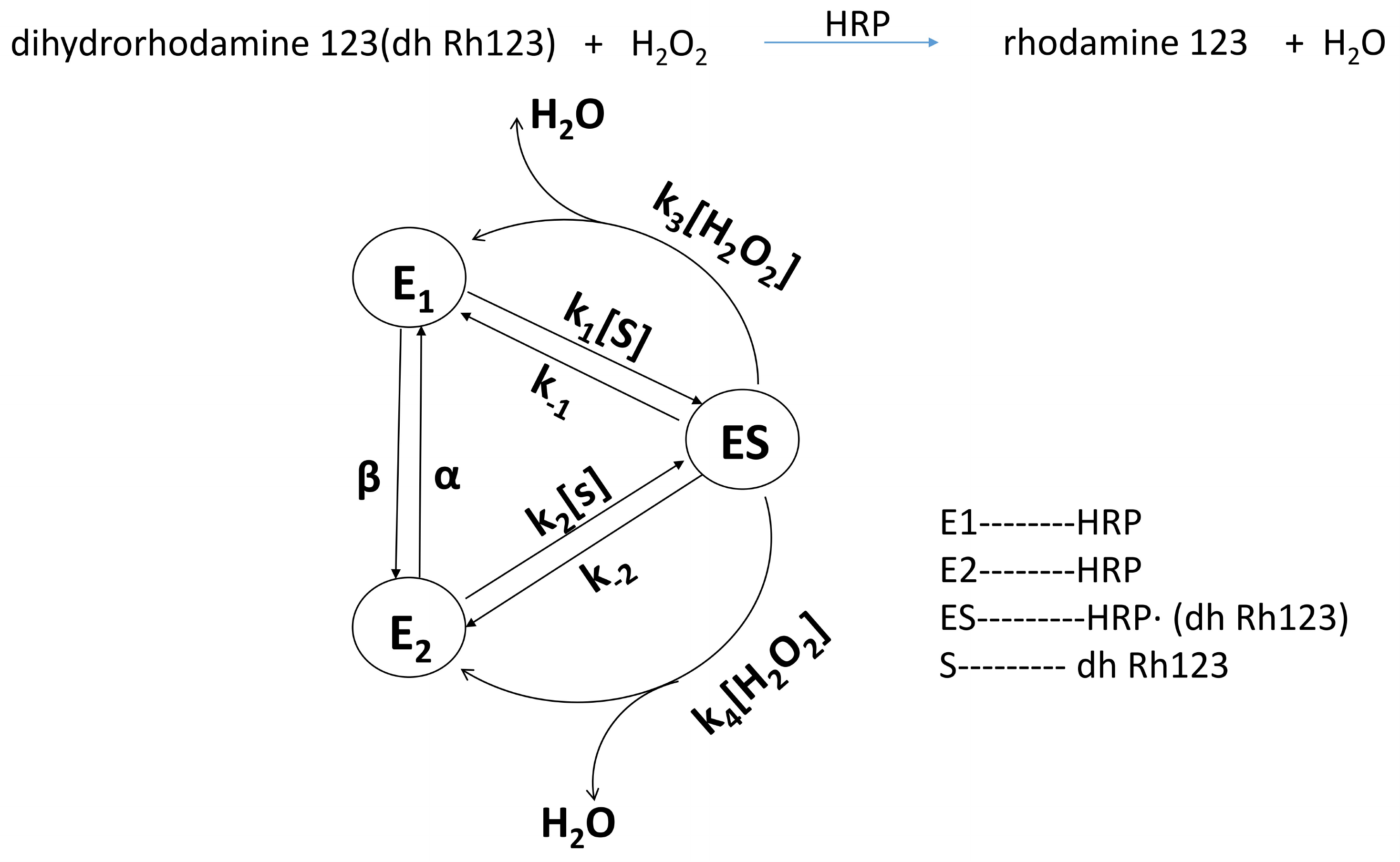}
\caption{The simplest kinetic scheme with two unbound enzyme states.}\label{fig02}
\end{figure}

The HRP can exhibit different reaction rates in different conformations
\cite{Rigler1, Rigler3, Rigler2}. A simplified reaction scheme includes the two conformations E$_1$ and E$_2$ in the reaction, representing the faster and slower parts. Since there is only one fluorescence molecule, that is the fluorophore rhodamine 123, the state of the complex is represented by ES. The enzyme molecule binds with substrate dihydrorhodamine 123 to form the ES complex while ES binds with substrate H$_2$O$_2$ and returns to the original enzyme by releasing H$_2$O molecule.

The simplest kinetic scheme shown in Figure. 2 includes conformational fluctuations in enzyme E. We assume that only ES has significant fluorescence signal. Thus, the stochastic kinetics of the single enzyme can be studied following the probability evolution of the states which is described by the master equation:

\begin{align*}
\frac{d}{{dt}}P &= \left[ {\begin{array}{*{20}{c}}
	{ - {k_1}[s] - \beta } & \alpha  & {{k_{ - 1}} + {k_3}[{H_2}{O_2}]}  \\
	\beta  & { - {k_2}[s] - \alpha } & {{k_{ - 2}} + {k_4}[{H_2}{O_2}]}  \\
	{{k_1}[s]} & {{k_2}[s]} & { - {k_{ - 1}} - {k_{ - 2}} - {k_3}[{H_2}{O_2}] -
		{k_4}[{H_2}{O_2}]}  \\
	\end{array}} \right]P \\
&= AP
\end{align*}
where \textbf{P }is a vector of [$P_{E_1}$, $P_{E_2}$, $P_{ES}$],
$P_{E_1}$, $P_{E_2}$, $P_{ES}$ are the probabilities of the enzyme being in the
$E_{1}$, $E_{2}$, ES states, respectively. A transition matrix \textbf{A} specifies the transition probability from one state to another.

\subsection*{Steady state flux and enzyme rate}

In equilibrium state, there is no net flux. However, for the non-equilibrium steady state, the net steady state flux is not necessarily zero, which can be defined as:

\begin{align*}
J&=P_{E_1}^SA(2,1)-P_{E_2}^SA(1,2)\\
&= \frac{{( - \alpha {k_1}{k_4}[{H_2}{O_2}] - \alpha {k_1}{k_{ - 2}} + \beta
		{k_2}{k_3}[{H_2}{O_2}] + \beta {k_2}{k_{ - 1}})[s]}}{{{\lambda _1}{\lambda _2}}}
\end{align*}
where the $\lambda{}$'s values can be obtained as the eigenvalues of the matrix
\textbf{A}. When there is no conformational change, i.e. $\alpha{}$=0, $\beta{}$=0,then J
=0. That is to say detailed balance can be preserved under no conformation changes. In addition, under conformational changes,
when $\alpha{}$ and $\beta{}$ are determined by
\begin{equation*}
\frac{\alpha }{\beta } = \frac{{{k_{ - 1}}{k_2} + {k_2}{k_3}[{H_2}{O_2}]}}{{{k_{
				- 2}}{k_1} + {k_1}{k_4}[{H_2}{O_2}]}}
\end{equation*}
the net flux is also zero, and the above formula is equivalent to the detailed
balance condition. Under conformational changes and without the above constraint, the net flux is non-zero, quantifying the degree of detailed balance breaking.

The transition rate matrix \textbf{A} governs the time evolution of the state probability and therefore acts as the driving force. It can be decomposed into two matrices as shown below (details given in the Supporting Information)\cite{Wang2008PNAS,Qian2002BPC}.

\begin{equation*}
A_{ij} =\left[ {\begin{array}{*{20}{c}}
	{A_{11}} & \frac{d_1}{{P_{2}^S}} & \frac{d_3}{{P_{3}^S}}\\
	\frac{d_1}{{P_{1}^S}} & {A_{22}} & \frac{d_2}{{P_{3}^S}}\\
	\frac{d_3}{{P_{1}^S}} & \frac{d_2}{{P_{2}^S}} & {A_{33}}\\
	\end{array}}\right]  
+\left[ {\begin{array}{*{20}{c}}
	{0} & 0 & \frac{J}{{P_{3}^S}}\\
	\frac{J}{{P_{1}^S}} & 0 & 0\\
	0 & \frac{J}{{P_{2}^S}} & 0\\
	\end{array}}\right]  
\end{equation*}

while $d_1=P_{2}^S A_{12}$, $d_2=P_{3}^S A_{23}$, $d_3=P_{1}^S A_{31}$

It can be checked that the above left rate matrix satisfies the detailed balance condition without net steady state flux, while the right rate matrix breaks the detailed balance explicitly with a net flux \textbf{J}. In other words, the driving force governing the probability evolution is determined by the detailed balance preserving equilibrium part and detailed balance breaking non-equilibrium part which is explicitly quantified by the non-zero steady state flux \textbf{J}. One can see that flux \textbf{J} itself has a rotational feature in state space from the above right rate matrix multiplied by its corresponding steady state probability. Therefore, the rotational flux serves as the non-equilibrium driving force.

The relationship between the steady-state turnover velocity or enzyme rate v and
the substrate concentration s becomes:
\begin{align*}
v &= P_{ES}^S({k_3} + {k_4})[{H_2}{O_2}]\\
&= \frac{{({k_3} + {k_4})[{H_2}{O_2}](\alpha {k_1} + \beta {k_2} +
		{k_1}{k_2}[s])[s]}}{{{\lambda _1}{\lambda _2}}}
\end{align*}

When the flux J=0,the equation is reduced to:

\begin{equation*}
v = \frac{{{k_1}{k_2}({k_3} + {k_4})[{H_2}{O_2}][s]}}{{{k_1}{k_2}[s] +
		({k_1}{k_4}[{H_2}{O_2}] + {k_2}{k_3}[{H_2}{O_2}] + {k_1}{k_{ - 2}} + {k_2}{k_{ -
				1}})}} \\
\end{equation*}
\begin{equation*}
or \qquad 1/v = {C_0} + {C_1}/[s]  \qquad  or \qquad 1/v = {C_0}' + {C_1}'/[{H_2}{O_2}]
\end{equation*}

The above two equations have MM relationship dependence on substrate concentrations [s] ([dh Rh123]) or [H$_{2}$O$_{2}$] for fixed
[H$_{2}$O$_{2}$] or fixed [s] ([dh Rh123]) respectively : The inverse of the enzyme rate is linear to the inverse of the substrate concentration. ${C_0}$ and ${C_1}$ as well as ${C_0}'$ and ${C_1}'$ are constants depending on the reaction parameters.

When the flux J$\not=$0, the equation is reduced to:
\begin{equation*}
v=\frac{(k_1k_2[s]^2+\alpha{}k_1[s]+\beta{}k_2[s])(k_3+k_4)[H_2O_2]}{k_1k_2[s]^2+F_1[s]+F_2}
\end{equation*}
\begin{equation*}
or \qquad 1/v = {C_0}'' + {C_1}''/[s] + {C_2}''/([s] + \lambda '') \\
\end{equation*}
\begin{equation*}
or \qquad 1/v={D_0} + {D_1}/[{H_2}{O_2}] + {D_2}/([{H_2}{O_2}] + \lambda_{D})
\end{equation*}
\begin{equation*}
where \qquad F_1=\alpha{}k_1+\beta{}k_2+k_1k_4[H_2O_2]+k_2k_3[H_2O_2]+k_1k_{-2}+k_2k_{-1}
\end{equation*}
\begin{equation*}
F_2=(\alpha+\beta)(k_3[H_2O_2]+k_4[H_2O_2]+k_{-1}+k_{-2})
\end{equation*}
Here, ${C_0}''$, ${C_1}''$, ${C_2}''$ and $\lambda ''$ as well as ${D_0}$, ${D_1}$, ${D_2}$ and $\lambda_{D}$ are constants depending on the
rate parameters for fixed substrate concentration of [H$_{2}$O$_{2}$] or [s] ([dh Rh123]) respectively. As seen the inverse of enzyme rate is no longer linear to the inverse of the substrate concentrations. The nonlinear dependence of the enzyme rate on the inverse of the substrate concentrations gives non-MM rate deviated from the conventional MM rate with linear dependence of inverse enzyme rate with respect to the substrate concentrations\cite{Xie2007, Cao,Qian2002BPC}. Therefore, the net flux breaking the detailed balance is the key here to determine whether the state is in equilibrium or non-equilibrium and whether the enzyme rate follows the Michaelis-Menton rate or not.

\subsection*{Autocorrelations and connections to flux and enzyme rate}

Because only the ES state has a fluorescence signal, we assume that the
fluorescence signal $f(i)$ for different enzyme states as:

\begin{equation*}
f(E_{1})=0, \quad f(E_{2})=0, \quad and \quad f(ES)=1
\end{equation*}

Then the mean fluorescence signal becomes

\begin{equation*}\label{key}
< f >  = f({E_1})P_{E_1}^S + f({E_2})P_{E_2}^S + f(ES)P_{ES}^S = P_{ES}^S
\end{equation*}

And the time correlation function becomes

\begin{equation*}
G(\tau ) =  < \Delta f\left( 0 \right)\Delta f(\tau ) >  = \sum\limits_{i,j \in
	\{ {E_1},{E_2},ES\} }^{} {f(i)f(j)} P_i^S{P_{ij}}(\tau ) -  < f{ > ^2}
\end{equation*}

\begin{equation*}
=P_{33}(\tau{}) P_{3}^{S} - (P_{3}^{S})^{2}
\end{equation*}

Due to the inevitable diffusion of some fluorescent molecules in the solution, the diffusion effect needs to be considered in the autocorrelations\cite{Rigler1}.

Therefore, the final form of the autocorrelation is given as
\begin{equation*}
G(\tau ) = {c_1}{e^{{\lambda _1}\tau }} + {c_2}{e^{{\lambda _2}\tau }} + {c_3} +
\frac{1}{N}{[1 + \frac{\tau }{{{\tau _D}}}]^{ - 1}}{[1 + \frac{\tau }{{{\tau
				_D}}}{\omega ^2}]^{ - 1/2}}
\end{equation*}
Here, $\tau{}$$_{D}$ is the diffusion rate depending on the diffusion coefficient. N is the mean number of particles in the effective observation volume, and $\omega$ is the aspect ratio of the radius and the depth of the laser point. After obtaining the experimental fluorescence data, we calculated the correlation function and then use above formula to fit the correlations and obtain the corresponding kinetic rate parameters. In this way, we can quantify the flux \textbf{J}, the enzyme rate $v$  and other related thermodynamic quantities such as entropy production rate (EPR) and chemical potentials.

The fluorescence signals and correlation plots are shown in Figure 3(a).

\begin{figure}[!ht]
\centering
\includegraphics[width=1.0\textwidth]{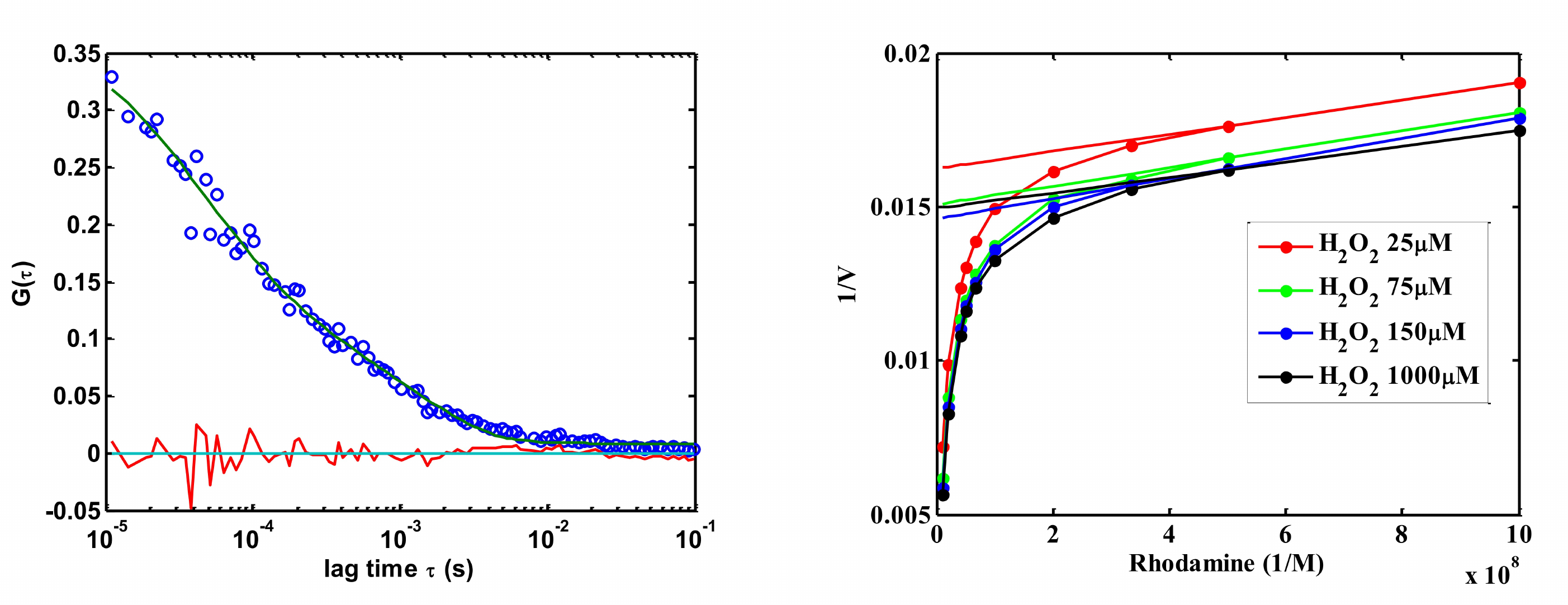}
\caption{(a) Fluorescence signals and autocorrelation. (b) Inverse enzyme rate versus inverse substrate concentration.}\label{fig03}
\end{figure}

\subsection*{Non-Michaelis-Menton Rate }

The non-Michaelis-Menton behavior of the enzyme rate with respect to the substrate is shown in Figure 3(b). The nonlinear relationship between the inverse enzyme rate and inverse substrate concentration is clearly demonstrated. The straight lines indicate the linear behavior which would have been expected for the Michaelis-Menton rate.

\subsection*{Flux quantification }

The quantified flux with respect to different substrate concentrations are shown in Figure 4(a). In the whole range of different substrate concentrations, the flux is non-zero. From the reaction scheme, we can see at relatively low concentrations of dihydrorhodamine 123, the flux increases with respect to the dihydrorhodamine 123. The increase of the substrate concentration of H$_{2}$O$_{2}$ also increases the flux at specific concentrations of dihydrorhodamine 123. The threshold in concentration of dihydrorhodamine 123 beyond which the flux drops or becomes flat depends on the H$_{2}$O$_{2}$ concentrations.

Based on the reaction scheme shown in Figure 1, let us assume that the steady state circular flux is given in the direction E$_1$ to ES to E$_2$, then the magnitude of the forward flux and the backward flux from E$_1$ to ES are given as P$_{E_1}$*k$_1$*[s] and P$_{ES}$*k$_3$*[H$_{2}$O$_{2}$]. The probabilities of E$_1$, E$_2$ and ES are denoted by P$_{E_1}$, P$_{E_2}$ and P$_{ES}$. According to the Equilibrium Law, changing the concentration of a chemical will shift the equilibrium to the side that would reduce that change in concentration. At certain H$_{2}$O$_{2}$ concentration, P$_{E_1}$+P$_{E_2}$ decreases as the concentration of dihydrorhodamine 123 increases and P$_{ES}$ increases accordingly. Also $P_{E_1}$+$P_{E_2}$ increases as the concentration of [H$_{2}$O$_{2}$] increases and $P_{ES}$ decreases accordingly at certain dihydrorhodamine 123 concentration.

At lower concentration of H$_{2}$O$_{2}$ and dihydrorhodamine 123, the forward flux and backward flux are both very small. When the dihydrorhodamine 123 concentration has a very small increase, the increase of forward flux is proportional to P$_{E_1}$*k$_1$, if the increase is very small, then the increase of forward flux is approximately proportional to forward flux. In the same way, the increase of backward flux is approximately proportional to the backward flux. This will lead to an increase of the net flux as the dihydrorhodamine 123 concentration increase, because the forward flux is larger than backward flux. As the concentration of dihydrorhodamine 123 increases even further, P$_{ES}$ will become even larger and P$_{E_1}$ will become smaller due to the Equilibrium Law, therefore, the increase of forward flux is less than the increase of backward flux, the net flux will decrease.

When the concentration of H$_{2}$O$_{2}$ increase, P$_{ES}$ will decrease, so the same addition of dihydrorhodamine 123 concentration will lead to the smaller addition of the P$_{ES}$, this reduces the increase of backward flux. Therefore, the net flux will also show the trend of increasing first and decreasing, but the decrease will be smaller.

At the higher concentration of H$_{2}$O$_{2}$, P$_{ES}$ will decrease even further. As the dihydrorhodamine 123 concentration increase, the flux will increase similar as that at the lower concentration of H$_{2}$O$_{2}$. As the dihydrorhodamine 123 concentration increase, P$_{E_1}$ will decrease and P$_{ES}$ will increase, the addition of forward and backward flux are proportional to P$_{E_1}$ and P$_{ES}$, respectively. The addition of forward flux will decrease, and the addition of backward flux will increase, so the net flux will keep on increasing slower and slower.

\subsection*{Chemical potential quantification}

The corresponding chemical potentials as the thermodynamic driving force can also be quantified accordingly based on the correlation data.

\begin{align*}
\Delta{}\mu{}&={k_B}T\ln
\frac{{{A_{31}}P_3^S{A_{12}}P_1^S{A_{23}}P_2^S}}{{{A_{13}}P_1^S{A_{32}}P_3^S{A_{21}}P_2^S}}\\
&={k_B}T\ln
\frac{{{A_{31}}{A_{12}}{A_{23}}}}{{{A_{13}}{A_{32}}{A_{21}}}}\\
&={k_B}T\ln \frac{{{d_1}{d_2}{d_3}}}{{(J + d_1)(J +{d_2})(J + {d_3})}}
\end{align*}
where ${d_1},{d_2},{d_3}$ are constants depending on the kinetic rate parameters. The chemical potentials at different substrate concentrations are shown in Figure 4(b). The chemical potentials are almost constants with respect to different rhodamine substrate concentrations at certain [H$_{2}$O$_{2}$]. The chemical potential acts as a voltage providing a quantification of the energy supply as a chemical battery.

\begin{figure}[!ht]
\centering
\includegraphics[width=1\textwidth]{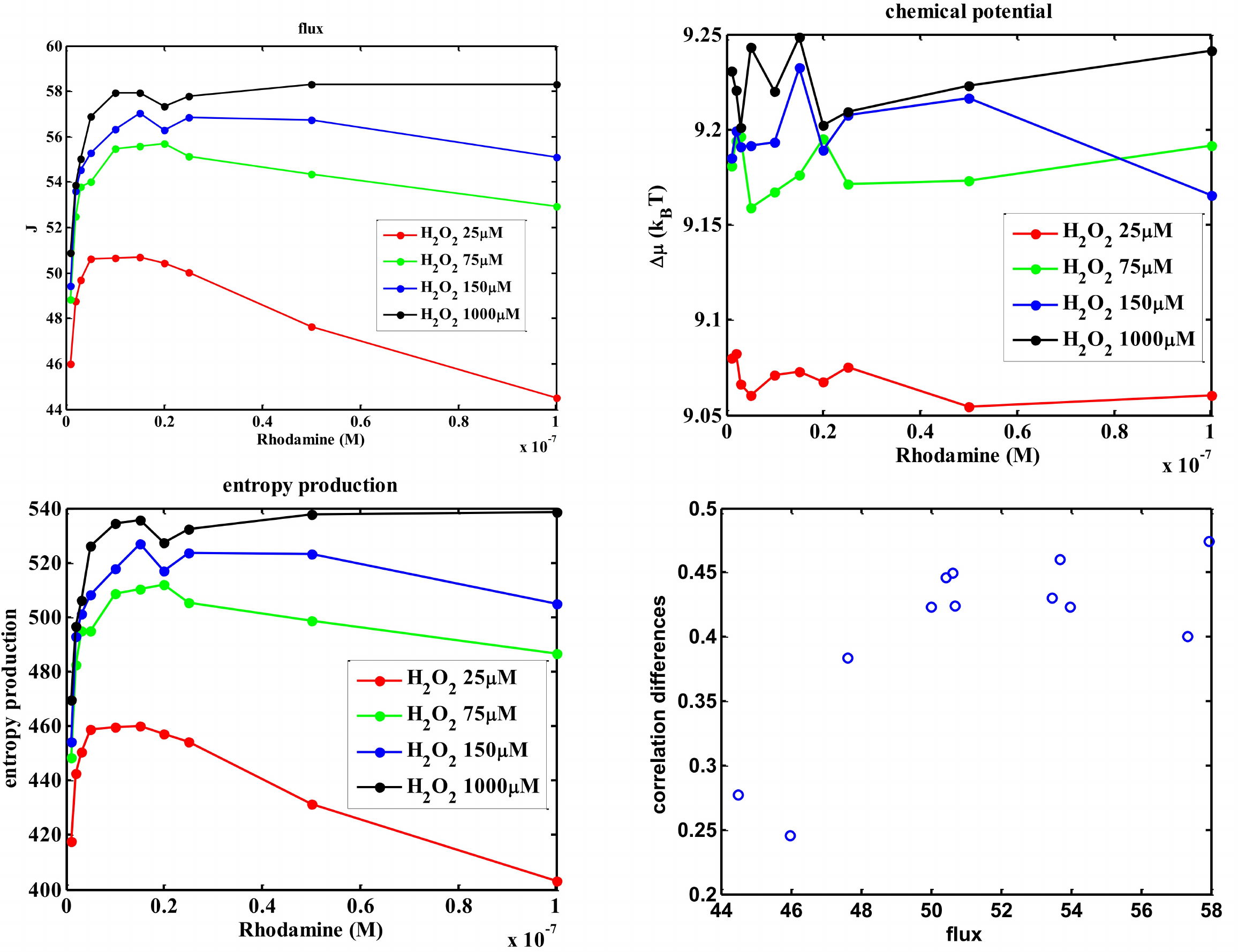}
\caption{(a) Flux values at different substrate concentrations. (b) Chemical potential at different substrate concentrations. (c) Entropy production rate at different substrate concentrations. (d) The correlation differences of the third-order autocorrelation function of forward and backward in time vs. the flux.}\label{fig04}
\end{figure}

\subsection*{Entropy production quantification}

The entropy production as a measure of the thermodynamic cost or dissipation for maintaining the non-equilibrium steady state can also be quantified according to the correlation data.
\begin{align*}
EPR &= \sum\limits_{i,j} {(P_i^S{A_{ji}} - P_j^S{A_{ij}})} \ln
\frac{{{A_{ji}}P_i^S}}{{{A_{ij}}P_j^S}}\\
&=J\ln \frac{{{A_{31}}{A_{12}}{A_{23}}}}{{{A_{13}}{A_{32}}{A_{21}}}}\\
&=J\ln \frac{{{d_1}{d_2}{d_3}}}{{(J + d_1)(J + {d_2})(J +
		{d_3})}}
\end{align*}

The ${d_1},{d_2},{d_3}$ are constants depending on the kinetic rate parameters. The entropy production rate at different rhodamine substrate concentrations are shown in Figure 4(c). The entropy production has similar behavior with respect to flux with respect to different substrate concentrations which can be easily understood since the flux is the origin of the non-equilibriumness and therefore the entropy production.

\subsection*{Quantification of time irreversibility }

In order to provide another measure of the non-equilibriumness of the single molecule enzyme reaction dynamics, we also calculate the higher-order autocorrelation functions from the experimental measurements. The third-order autocorrelation function is shown as follows

\begin{equation*}
G({\tau _1},{\tau _2}) =  < I(0)I({\tau _1})I({\tau _1} + {\tau _2}) >
\end{equation*}
where I(0), I($\tau{}$$_{1}$) and I($\tau{}$$_{1}$+$\tau{}$$_{2}$), are the
intensity of the fluorescence at time 0,$\tau{}$$_{1}$,
and$\tau{}$$_{1}$+$\tau{}$$_{2}$. When the enzyme reaction is in equilibrium and with no net flux, its third-order correlation function is symmetric in forward and backward time directions. However, an experimental observation of non-symmetric G($\tau{}$$_{1}$,$\tau{}$$_{2}$) in forward and backward direction in time will be an indication for time irreversibility and therefore the non-equilibriumness\cite{ Li2011BJ,Qian2004PNAS}.

We calculated the third-order correlation function from the experimental data, the corresponding result is a two-dimensional matrix, normalized by the maximum value. The difference between the matrix and its transpose representing the asymmetry can be calculated, which is also the correlation difference forward and backward in time. In order to show the correlation difference clearly, we take the absolute maximum value of the difference matrix for each data. The results show in Figure 4(d) that, as the flux increases, the correlation differences of the third-order autocorrelation in forward and backward in time direction increases. In other words, as the flux increases, the irreversibility in time increases.

\subsection*{The quantifications of the origin of the flux, chemical potential and
entropy production as the heat absorbed in enzyme reaction }

In order to uncover the physical origin of the underlying chemical potential, we quantify the corresponding Gibbs free energy change of the enzyme reaction through Isothermal titration calorimetry(ITC) measurements.

According to enzyme reaction scheme Figure 2, there are two main paths of the enzyme reaction. The corresponding state changes are E$_{1}$-ES-E$_{2}$, and
E$_{2}$-ES-E$_{1}$. Or one path starts from E$_{1}$, the other starts from
E$_{2}$. According to the fitting of FCS correlations, the steady-state
probabilities of E$_{1}$ and E$_{2}$ are 0.09 and 0.91, respectively. In
Isothermal titration calorimetry (ITC) measurements\cite{ITC},
E$_{1}$ and E$_{2}$ are equivalent to two relatively independent sets. The enthalpy change and binding constants of the underlying enzyme reaction can be obtained by fitting to the heat measurements. The difference between the Gibbs free energy of
the two paths should correspond to the chemical potential driving the net flux that is fitted by FCS. $\Delta{}$G=$\Delta{}$G$_2$-$\Delta{}$G$_1$= 28.37 KJ/mol. The Gibbs free energy change is therefore 28.37 KJ/mol. The Gibbs free energy change is very close to the chemical potential fitted by the FCS data 23.5 KJ/mol(9.25 kT). This illustrates that the origin of the chemical potential or the flux is the heat
absorbed in the enzymatic reaction.

\section*{Conclusions: }

In this work, we have experimentally explored the single molecule enzyme reaction dynamics. The enzyme reaction rate shows significant deviations from the conventional Michaelis-Menton behavior. We found that the non-equilibrium flux is the driving force for the non-MM behavior. We quantified this non-equilibrium flux for the dynamics and the corresponding chemical potential as well as the entropy production rate for the non-equilibrium thermodynamics. We uncovered the origin of the non-equilibrium flux, the chemical potential and entropy production rate for driving enzyme reaction as the heat absorbed (energy input) in the enzyme reaction. We also quantified the time irreversibility for further characterizing the non-equilibrium behavior. Our work provides a framework for quantifying the non-equilibrium signature both at the dynamic and the thermodynamic level. We expect such framework can be applied to other interesting non-equilibrium physical and biological systems to quantify the driving force for the non-equilibrium dynamics and the underlying thermodynamics.

\section*{Acknowledgement}
This work was support in part by National Natural Science Foundation of China ( NSFC-91430217) and NSF-PHY-76066 (USA).



\begin{thebibliography}{10}

\bibitem{Jackson} Jackson, E.A.Perspectives of
Nonlinear Dynamics.Cambridge Unversity Press, Cambridge. 1989

\bibitem{Wang2008PNAS} J. Wang, L. Xu, E. K.
Wang. Potential landscape and flux framework of nonequilibrium networks:
robustness, dissipation, and coherence of biochemical oscillations. Proc. Natl.
Acad. Sci. USA , 105: 12271-12276. (2008).

\bibitem{Wang2015AP} J. Wang*, ~Landscape and
flux theory of non-equilibrium dynamical systems with application to biology,
Advances in Physics, 64:1, 1-137. (2015).

\bibitem{Xie2006} English, Brian P.; Min, Wei;
van Oijen, Antoine M.; Lee, Kang Taek; Luo, Guobin; Sun, Hongye; Cherayil, Binny
J.; Kou, S.C.; Xie, X. Sunney
"Ever-fluct\href{https://bernstein.harvard.edu/papers/brian\_nchembio\_2006.pdf}{uating
	single enzyme molecules: Michaelis-Menten equation revisited,"~\textit{Nat.
		Ch}}\textit{em. Bio}.,~\textbf{2}, 87 (2006).

\bibitem{Xie2013}Xie, X. Sunney. "Enzyme
Kin\href{https://bernstein.harvard.edu/papers/Science-2013-Xie-1457-9.pdf}{etics,
	Past and Present,"~\textit{Science}}~\textbf{342}, 1457-1459
DOI:10.1126/science.1248859 (2013)

\bibitem{MichaelisMenton} Michaelis, L. and
Menten, M.L.Die Kinetik der Invertinwirkung. Biochem Z.49,333-369 (1913).

\bibitem{Moerner} ~W. E. Moerner and D. P. Fromm,
''Methods of Single-Molecule Fluorescence Spectroscopy and
Microscopy,''~\textbf{invited review},~\textit{Rev. Sci. Instrum}.~\textbf{74},~
3597-3619 (2003).

\bibitem{Xie2007} Min, Wei; Gopich, Irina V.;
English, Brian P.; Kou, Sam C.; Xie, X. Sunney.; Szabo, Attila "When Does
\href{https://bernstein.harvard.edu/papers/weimin\_jpcb\_2006.pdf}{the
	Michaelis-Menten Equation Hold for Fluctuating Enzymes?"~\textit{J.
		Phys.}}\textit{ Chem.~}B,~\textbf{110}, 20093-7 (2006).

\bibitem{Cao}Jianshu Cao. MichaelisMenten
Equation and Detailed Balance in Enzymatic Networks. J. Phys. Chem. B. 115, 5493--5498 (2011)

\bibitem{Qian2002BPC} Hong Qian , Elliot L.
Elson. Single-molecule enzymology: stochastic Michaelis--MentenKinetics.
Biophysical Chemistry 101 -- 102, 565--576(2002)


\bibitem{Rigler1} Lars  Edman,  Zeno
FiSldes-Papp,  Stefan Wennmalm,  Rudolf Rigler, The  fluctuating  enzyme:  a
single  molecule  approach. Chemical Physics 247 ,11-22 (1999).

\bibitem{Rigler3} Lars Edman and Rudolf Rigler. Memory landscapes of single-enzyme molecules. Proc. Natl. Acad. Sci. 97, 8266-8271 (2000).

\bibitem{Rigler2}Kai Hassler, Per Rigler,, Hans
Blom,, Rudolf Rigler, Jerker Widengren, and Theo Lasser. Dynamic disorder in
horseradish peroxidase observed with total internal reflection fluorescence
correlation spectroscopy. OPTICS EXPRESS. 15,  5366-5375(2007)

\bibitem{Li2011BJ} Chunhe Li, Erkang Wang, and
Jin Wang. Landscape, Flux, Correlation, Resonance, Coherence, Stability, and Key
Network Wirings of Stochastic Circadian Oscillation. Biophysical Journal 101,
1335--1344 (2011)

\bibitem{Qian2004PNAS}Qian, H., and E. L.
Elson.Fluorescence correlation spectroscopy with high-order and dual-color
correlation to probe nonequilibrium steadystates.Proc. Natl. Acad. Sci. USA.101:2828--2833 (2004).

\bibitem{ITC} O'Brien, R.,
Ladbury, J.E. and Chowdry B.Z. Isothermal titration calorimetry of biomolecules.
Chapter 10 in Protein-Ligand interactions: hydrodynamics and calorimetry Ed.
Harding, S.E. and Chowdry, B.Z, Oxford University Press. (2000)

\end{thebibliography}
\end{document}